\documentclass[%
 reprint,
 amsmath,amssymb,
 aps,
]{revtex4-2}

\usepackage{tikz}
\usetikzlibrary{decorations.pathmorphing,patterns}
\usepackage{tikzit}

\tikzstyle{motor}=[fill=red, draw=red, shape=circle, scale=0.5]
\tikzstyle{cargo}=[shape=circle, fill=blue, draw=blue]

\tikzstyle{filament}=[->, draw=green, ultra thick]
\tikzstyle{undirected filament}=[-, draw=green, ultra thick]
\tikzstyle{cell membrane}=[-, draw=yellow]
\tikzstyle{spring}=[-, decoration={{aspect=0.3, segment length=0.5mm, amplitude=0.5mm,coil}}, draw=black, decorate]

\usepackage{pgfplots}
\pgfplotsset{compat=1.17} 
\usepgfplotslibrary{dateplot}

\usepackage{orcidlink}   
\usepackage{hyperref}    

\usepackage{graphicx}
\usepackage{dcolumn}
\usepackage{bm}

\begin{document}

\preprint{APS/123-QED}

\title{Applying a Gaussian  networking theory \\ to model motor-driven transport\\
along cytoskeletal filaments}

\author{Nadine du Toit \orcidlink{0000-0003-0262-7010}}
\email{Corresponding author: 24461989@sun.ac.za}
\affiliation{
Department of Physics, Stellenbosch University, Stellenbosch 7602, South  Africa\\
}
 
\author{Kristian K. M\"uller-Nedebock \orcidlink{0000-0002-1772-1504}}
 \email{kkmn@sun.ac.za}
\affiliation{
Department of Physics, Stellenbosch University, Stellenbosch 7602, South  Africa
}
\affiliation{National Institute for Theoretical and Computational Sciences, Stellenbosch, 7602, South Africa}




\date{\today}

\begin{abstract}
This paper builds on a recently introduced dynamical networking framework, applying it to model motor-driven transport along cytoskeletal filament networks. Within this approach, the networking functional describes the periodic binding and unbinding of motors to available filament sites,whilst accounting for all possible pairing, enabling a field-theoretic treatment of constrained motion in complex networks. In this application, the dynamical networking theory is introduced into a Martin-Siggia-Rose representation of the Langevin dynamics describing the motion of a motor protein and its cargo. Results are presented in a collective description of motors on a network, for two different scenarios, namely homogeneous and non-homogeneous networks. A diffusion coefficient is presented for homogeneous networks, whilst it is shown that various possibilities remain for disordered averaging over network densities for non-homogeneous networks. 
\end{abstract}

\maketitle

\section{\label{sec:intro}Introduction}

In active matter, one typically encounters various proteins which, depending on the context, are able to form 
bonds or reversibly associate with 
with other types of particles, spatially constraining or cross-linking the particles to one another. In this work, we apply a recently developed field theoretical framework \cite{dutoitDynamicalNetworkingUsing2025} that allows one to impose 
linking 
constraints at specified time intervals. Here we apply this framework by to model molecular machines transporting cargoes along cytoskeletal filaments. The method could also be extended to describing various other intracellular processes.
In the actin cytoskeleton, for example, one actin filament is able to bind to another with the aid of the Arp2/3 complex \cite{Guerin2010}. In the context of motor-driven transport, cytoskeletal motor proteins, propel themselves along the length of the filaments of the cytoskeleton whilst dragging along various vesicles and organelles. The heads of motor proteins, such as kinesins, consist of two arms that, in the presence of ATP, bind and unbind to the cytoskeleton filaments, in order to 
move along in a step-like motion \cite{nelsonBiologicalPhysicsEnergy2020}. As part of the same process, the tails of the motor proteins are able to bind to various cargoes, such that they may be dragged towards other regions of the cell, after which they may unbind again. These examples illustrate the diversity of dynamical scenarios in which intracellular particles are spatially 
associated or linked with one another. 
Motor proteins, in particular, provide a natural setting for applying the dynamical networking framework, since their step-like motion arises from periodic binding and unbinding events along cytoskeletal filaments. Importantly, such transport is intrinsically an active process, driven by the consumption of chemical energy. Furthermore, how these active binding–unbinding events collectively affect behaviour on the scale of the entire cell remains an ongoing research effort \cite{Burute2019, Guerin2010}.

Motor proteins and  filaments of the cytoskeleton have been modelled through various theoretical and mathematical perspectives, some focusing on specific types of motor proteins \cite{Trotta}, whilst others provide models of generic motors \cite{Julicher1997}. Computational calculations and molecular dynamics simulations of motor-driven transport have also lead to significant insights\cite[see e.g., ][]{Jung2019}. Coarse-grained models highlight  that the spatial organisation of the cytoskeleton affects intracellular transport \cite{Hafner2018} and targeted delivery \cite{Hafner2016} of cargoes. The exact role that the organisation of the cytoskeleton plays in specific aspects of this transport process, however, remains an active area of research \cite[see e.g.,][]{Burute2019}. 

Sophisticated imaging techniques allow the investigation of vesicle dynamics within complex cellular contexts and processes \cite[see e.g., ][]{DuToit2018}. Live cell fluorescence microscopy and single particle tracking have allowed the measurement of speeds at which specific cargoes are transported within the cell \cite{Bandyopadhyay2014}. High resolution single-molecule microscopy has  also lead to significant insights into the various types of motor proteins \cite{Vale2003} and the mechanisms by which they move \cite[see e.g.,][]{DyneinPaper}. 

Whilst motor proteins have been studied from various perspectives, current theoretical models tend to be insufficient to model the complexities seen in vivo, even when considering only a single motor \cite{SMogre2020}. Modelling the collective behaviour of motor proteins poses additional challenges. Collective models can be useful, for example, for accounting for the effects of the hydrodynamic coupling of motors to one another \cite{Guerin2010}. Various models of the collective behaviour of motor proteins and filaments in contexts other than motor-driven transport have also been developed  \cite{JulicherProst1995,Banerjee2011}. 

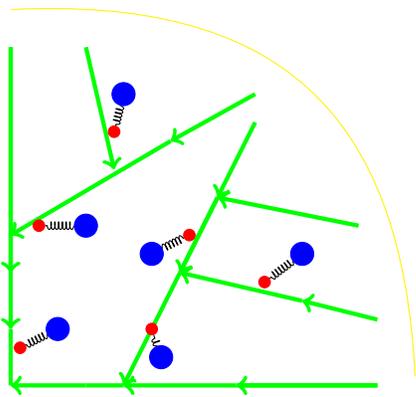
\begin{figure}
    \centering
\begin{tikzpicture}[scale = 0.5]
	\begin{pgfonlayer}{nodelayer}
		\node [style=none] (0) at (0, 0) {};
		\node [style=none] (1) at (2, 0) {};
		\node [style=none] (5) at (3, 0) {};
		\node [style=none] (12) at (4.5, 3) {};
		\node [style=none] (13) at (6, 0) {};
		\node [style=none] (14) at (5, 0) {};
		\node [style=none] (33) at (9.75, 0) {};
		\node [style=none] (34) at (5.5, 5) {};
		\node [style=none] (36) at (7.75, 2.25) {};
		\node [style=none] (42) at (9.75, 1.75) {};
		\node [style=none] (43) at (9.25, 4.25) {};
		\node [style=none] (44) at (6.5, 7) {};
		\node [style=motor] (45) at (3.75, 1.5) {};
		\node [style=motor] (46) at (6.75, 2.75) {};
		\node [style=none] (48) at (0, 0) {};
		\node [style=none] (49) at (0, 1.5) {};
		\node [style=none] (51) at (0, 3) {};
		\node [style=none] (52) at (0, 4) {};
		\node [style=none] (53) at (0, 9) {};
		\node [style=none] (55) at (6, 2.75) {};
		\node [style=none] (57) at (4.25, 6.5) {};
		\node [style=motor] (63) at (0.25, 1) {};
		\node [style=none] (65) at (6.5, 7.75) {};
		\node [style=none] (66) at (4.25, 6.5) {};
		\node [style=cargo] (68) at (4, 0.75) {};
		\node [style=cargo] (69) at (7.75, 3.5) {};
		\node [style=cargo] (70) at (1.25, 1.5) {};
		\node [style=none] (72) at (0, 0) {};
		\node [style=motor] (89) at (0.75, 4.25) {};
		\node [style=none] (90) at (0, 0) {};
		\node [style=motor] (105) at (4.75, 4) {};
		\node [style=motor] (106) at (2.75, 6.75) {};
		\node [style=cargo] (110) at (2, 4.25) {};
		\node [style=cargo] (111) at (3.75, 3.5) {};
		\node [style=cargo] (112) at (3, 7.75) {};
		\node [style=none] (116) at (10.75, 0.25) {};
		\node [style=none] (117) at (0, 10) {};
		\node [style=none] (120) at (0, 6) {};
		\node [style=none] (122) at (2, 9) {};
		\node [style=none] (123) at (2.75, 5.75) {};
	\end{pgfonlayer}
	\begin{pgfonlayer}{edgelayer}
		\draw [style=filament] (1.center) to (0.center);
		\draw [style=undirected filament] (5.center) to (1.center);
		\draw [style=filament] (12.center) to (5.center);
		\draw [style=filament] (13.center) to (5.center);
		\draw [style=undirected filament] (34.center) to (12.center);
		\draw [style=undirected filament] (33.center) to (13.center);
		\draw [style=filament] (36.center) to (12.center);
		\draw [style=filament] (33.center) to (13.center);
		\draw [style=filament] (34.center) to (12.center);
		\draw [style=filament] (44.center) to (34.center);
		\draw [style=filament] (43.center) to (34.center);
		\draw [style=filament] (42.center) to (36.center);
		\draw [style=undirected filament] (49.center) to (48.center);
		\draw [style=filament] (51.center) to (49.center);
		\draw [style=undirected filament] (53.center) to (51.center);
		\draw [style=filament] (57.center) to (52.center);
		\draw [style=filament] (53.center) to (51.center);
		\draw [style=filament] (65.center) to (66.center);
		\draw [style=cell membrane, bend left=45, looseness=1.25] (117.center) to (116.center);
		\draw [style=spring] (110) to (89);
		\draw [style=spring] (111) to (105);
		\draw [style=spring] (112) to (106);
		\draw [style=spring] (69) to (46);
		\draw [style=spring] (68) to (45);
		\draw [style=spring] (70) to (63);
		\draw [style=filament] (122.center) to (123.center);
	\end{pgfonlayer}
\end{tikzpicture}
    \caption[Schematic diagram of motor proteins on a cytoskeleton]{Schematic diagram of generic motor proteins (red) transporting cargoes (blue) on the branched filaments (green) of the cytoskeleton within a cell. A portion of the cell membrane is shown in orange.}
    \label{fig:transportInCell}
\end{figure}

Despite extensive studies of individual motors, a unified theoretical description of their collective transport across different cytoskeletal network architectures and cellular environments remains absent. Such a model would have to account for the directed diffusion of motor proteins along a filamentous network, as depicted in Fig. 
\ref{fig:transportInCell}. This poses the challenge of constraining the motion of the motor proteins to some spatial configuration of a filament network, which may include various branches and intersections. 

The  formulation of motor-driven transport presented here, builds directly on the dynamical Gaussian networking framework developed in Ref.~\cite{dutoitDynamicalNetworkingUsing2025}, which we combine with a Martin-Siggia-Rose representation of the Langevin  dynamics  of a motor and its cargo. The networking  theory plays the essential role of providing a mathematical mechanism by which one can periodically attach a motor protein to one of a set of  possible sites distributed along the length of a cytoskeleton filament, whilst the Langevin dynamics describe the diffusion of the motor proteins and their cargoes within the cytoplasm. The novel aspect of the present work is that motor transport is treated as an intrinsically active process: rather than being purely diffusive, motor motion is modelled with a constant bias or effective speed that captures the directed transport arising from underlying energy consumption. By utilising the networking theory in  this manner, the formalism presents the opportunity to model the directed diffusion of motors and their cargoes along various configurations of cytoskeletal filaments, including branches and intersections, potentially even allowing one to account for spatial and/or temporal filament fluctuations via disordered averaging.

 To start off the discussion, Section~\ref{sec:DynamicalMotorCargoCoupling} introduces the Langevin and Martin–Siggia–Rose formulation for the coupled dynamics of a motor and its cargo. 
In Section~\ref{sec:networking}, the dynamical networking functional is applied to describe the stochastic attachment of the motor to filament binding sites. 
Section~\ref{sec:collective} develops a collective description of motors on a network, including small fluctuation expansions. 
The main results are presented in Section~\ref{sec:results}, where transport is analysed for both homogeneous and non-homogeneous filament networks. 
Conclusions are given in Section~\ref{sec:Conclusions}, while technical details, including the random phase approximation and the saddle point approximation, are provided in the appendices.

\section{Dynamical motor and  cargo coupling}
\label{sec:DynamicalMotorCargoCoupling} 
It is assumed that the connection of the cargo to the motor is elastic, such that the Hamiltonian is given by the elastic potential energy of this connection. Assuming that this connection has a spring constant $\kappa$\,, the Hamiltonian is then given by:

\begin{equation}
H = \frac{\kappa}{2} (R (t) - x(t) )^2
\label{eq:HamiltonianStraight}
\end{equation}

\begin{figure}[ht]
\begin{tikzpicture}

\draw[decoration={aspect=0.3, segment length=2mm, amplitude=1mm,coil},decorate] (4,0.2) -- (2,1.2);

 \draw[line width=5pt,green][->] (0,0) -- (5,0) ;
 \draw[line width=5pt,green] (4.8,0) -- (7,0) ;

\draw[red,fill=red] (4,0.2) circle (1ex);
\draw[thick][->] (0,0.0) -- (4,0.0) node at (4,-0.3) {\small \color{red}$x(t) $};


\draw[blue,fill=blue] (2,1.2) circle (3ex) ;
\draw[thick][->] (0,-0.10) -- (2,-0.10) node at (2, -0.35) {\color{blue} $R(t)$};

\node [style=none] at (4.75,1) {\normalsize  spring with coefficient $\kappa$};
\node [style=none] at (1, 1.25) {\normalsize \color{blue} cargo};
\node [style=none] at (4, -0.5) {\normalsize \color{red} motor protein};
\end{tikzpicture}
\label{fig:straightFilamentDiagram}
\caption{Schematic diagram depicting the coupling of a cargo with position $R(t)$ to a motor protein with position $x(t)$.}
\end{figure}
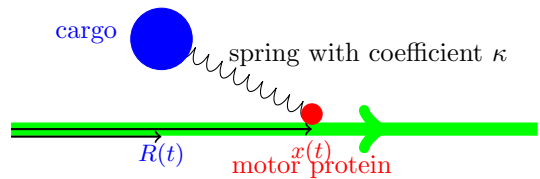 

It should be noted that the equations above make use of the scalars $x(t)$ and $R(t)$, since only the motion along the direction of motion of the motor protein is considered (see Fig. \ref{fig:straightFilamentDiagram}). This has been done to simplify the mathematics and will provide all the necessary information for the positions, speed and diffusion of the motor and cargo. 
The model of this system can be set up using two coupled Langevin equations; one for the position of the motor $x(t)$ and one for the position of its cargo $R(t)$:
\begin{subequations}
\label{eq:LangevinEqns}
\begin{eqnarray}
\mathcal{L}_x = - \gamma_x \dot{x} (t) -\frac{\partial H}{\partial x (t)} + f_\mathrm{drift} + f_x (t) =0 \, , \\
\mathcal{L}_R = - \gamma_R \dot{R} (t) -\frac{\partial H}{\partial R (t)} + f_R (t)=0\, .
\end{eqnarray}
\end{subequations}
Here,$\gamma_x$ and $\gamma_R$ are the drag coefficients of the motor and cargo, respectively. The forces $f_x(t)$ and $f_R (t)$ are stochastic forces acting on the motor and cargo. These forces are included to account for the effects of thermal noise and are thus defined to be Gaussian correlated. The parameters $\lambda_R$ and $\lambda_x$ can be interpreted as a measure of strength of each of the stochastic forces and adhere to the fluctuation dissipation theorem such that
\begin{eqnarray}
\lambda_{x,R} ={}& 2 \gamma_{x,R} k_\mathrm{B} T\,.    \label{eq:lambdax}
\end{eqnarray}

Finally, it is assumed that the motor propels itself forward at a drift speed $v_\mathrm{drift}$ due to a constant driving force $f_\mathrm{drift} = \gamma_x v_\mathrm{drift}$ acting on the motor. It may be worth pointing out that although the magnitude of this driving force is assumed to be constant in time, it may vary depending on the load of the cargo. This model, is however simplified by assuming that the motor is to be attached to the same cargo for the duration of its motion.

Fig. \ref{fig:straightFilamentDiagram} also depicts a directed filament, which does not explicitly form part of these Langevin equations. This will be included via a dynamical networking functional, such that the motor protein is required to move along the length of some filament configuration. The benefit of modelling the attachment of the motor to a filament in this way, is that the model allows for the consideration of not only single filaments, but also networks of branched and intersecting filaments. Since the focus of this discussion is on the dynamical implementation of the networking functional to the motion of the motor, a one-dimensional representation of the motor protein's motion is sufficient. Here, the drift speed of the motor has merely been included in the Langevin equation. If one were to extend the model to more dimensions so as to include branches and intersections, the orientation of the filament would need to be accounted for when assigning a drift speed to the motor. 

In preparation for the introduction of a networking functional, the coupled Langevin equations (eqs.~\eqref{eq:LangevinEqns}) may be rewritten, in the Martin-Siggia-Rose (MSR) formalism, using a \textit{generating functional}, as follows:
\begin{widetext}
\begin{equation}
Z[J(t)] =\mathcal{N}\int [\mathrm{d} x(t)]  [\mathrm{d} \hat{x}(t)]  [\mathrm{d} R(t)][\mathrm{d} \hat{R}(t)] [\mathrm{d} f_x(t)] [\mathrm{d} f_R(t)] \mathrm{e}^{\mathrm{i} \int_t \hat{x}(t) \mathcal{L}_x+ \mathrm{i} \int_t \hat{R}(t) \mathcal{L}_R-\frac{1}{2 \lambda_x} \int_t f_x^2(t)-\frac{1}{2 \lambda_R} \int_t f_R^2(t) +\int_t J(t) x(t)}.
\label{eq:Zraw}
\end{equation}
\end{widetext}
The interested reader is urged to consult \cite{Jensen1981a} and \cite{Jouvet1979} for a thorough account of the MSR  formalism. In the expression above, $\mathcal{N}$ is a normalisation constant. Here, the square braces indicate path integrals whilst the shorthand $\int _t$ has been used to indicate integrals over time, i.e. $\int _{-\infty} ^\infty  \mathrm{d}t $. This generating functional  (eq.~\eqref{eq:Zraw}) contains all of the information of the dynamics of the motor and the cargo including all possible realisations of the stochastic forces. The fields $\hat{x}$ and $\hat{R}$ are known as \textit{response fields} or \textit{auxiliary fields} \cite{Jouvet1979}. These fields couple to the Langevin equations and account for the dynamics of the fields $x(t)$ and $R(t)$ respectively. The response fields are required to obey a set of causality rules \cite{Jensen1981a} which ensure that the discretisation of the time-dependent path integrals remains causal. In addition to the Langevin equations, eq.~\eqref{eq:Zraw} also contains the Gaussian probability distributions of both stochastic forces i.e. $f_x$ and $f_R$. Implementing the functional integrals over the stochastic forces together with these probability distributions is equivalent to taking averages over all possible realisations of $f_x(t)$ and $f_R(t)$.  Finally, a \textit{source term}, $J(t)$ is coupled to $x(t)$ in eq.~\eqref{eq:Zraw}. This term may be used to calculate averages and correlation functions for the position of the motor as follows:
\begin{equation}
\langle x(t_1) x(t_2).... x(t_n) \rangle = \frac{1}{Z[J(t)]}\left. \frac{\partial^n Z[J(t)]}{\partial J(t_1)\partial J(t_2)....\partial J(t_n)}    \right|_{J=0}\,.
\label{eq:aveFormula}
\end{equation}

In order to obtain such averages, however, one first needs to evaluate the path integrals in eq.~\eqref{eq:Zraw}. Tools and approximation schemes for implementing such integrals analytically are readily available \cite[see e.g., ][]{Fantoni2011, Mateyisi2014}. 

\section{Networking a motor to a filament}
\label{sec:networking}

As it stands, the generating functional (eq.~\eqref{eq:Zraw}) does not account for the filament configuration along which the motor may walk, i.e. which sites it may visit. This can be remedied by incorporating a meticulously designed networking functional, following Ref.~\cite{dutoitDynamicalNetworkingUsing2025}. To do this, the filament configuration will be represented as a discrete set of positions. These positions, from here onward referred to as \textit{binding sites}, will act as possible attachment points for a motor protein. 

The networking functional needs to count all possible configurations in which the position $x(t)$ of a single motor protein can coincide with one of  $n$ possible binding sites with positions $r_1,r_2,r_3,..,r_n$. A networking functional needs to be constructed such that it  allows the joining of the position of a single motor to multiple possible points along the filament. Following the detailed derivation in Ref.~\cite{dutoitDynamicalNetworkingUsing2025}, a suitable networking functional for the position of the motor at a time $t_*$ may be composed by using the following Gaussian integrals: 
\begin{subequations}
\begin{eqnarray}
\int [ \mathrm{d}\Phi] [\mathrm{d} \Phi^*]\, \Phi(r,t)\, \mathrm{e}^{-\alpha \int_{y,t} \, \Phi(y,t)\,\Phi^*(y,t)}= 0\,,  \\
\int [ \mathrm{d}\Phi] [\mathrm{d} \Phi^*]\, \Phi^*(r,t) \, \mathrm{e}^{-\alpha \int_{y,t} \, \Phi(y,t)\,\Phi^*(y,t)}=0\,.
\end{eqnarray}
\label{eq:networking}
\end{subequations}
Manipulating the combinations of these Gaussian integrals, one may obtain:
\begin{multline}
Q[x(t_*),t_*]=\mathcal{N}_\Phi \int [ \mathrm{d}\Phi] [\mathrm{d} \Phi^*]\,\,  \Pi_n(1+\Phi(r_n,t))\,\\
\times\Phi^*(x(t_*),t_*) \mathrm{e}^{- \alpha \int_{y,t} \, \Phi(y,t)\,\Phi^*(y,t)}    \, .     
\label{eq:networkingConstraintRaw}    
\end{multline}
such that, eqs.~\eqref{eq:networking} reveal that many of the terms vanish upon evaluation of the functional integrals. This leaves
\begin{equation}
Q[x(t_*),t_*]=\frac{1}{\alpha} \Sigma_n \delta(r_n - x(t_*))\delta(t - t_*)\,,
\end{equation}
which counts all the possible ways that a motor may attach to one of the $n$ binding sites at a given time $t^*$, as desired. 

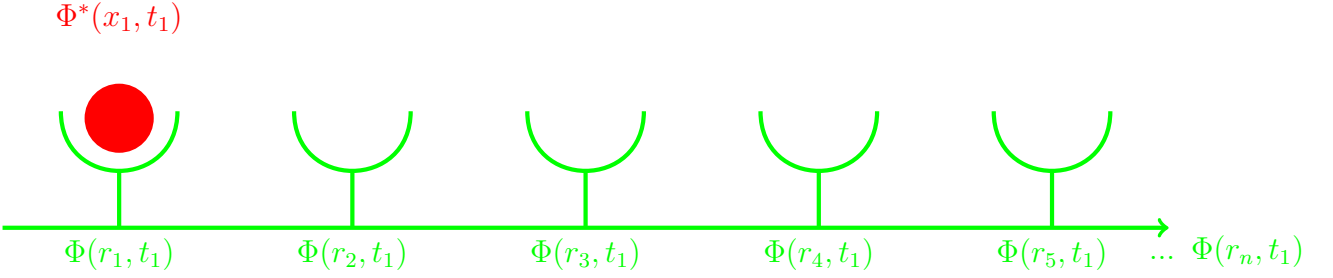
\begin{figure*}
\begin{tikzpicture}[scale=1.55]
		\node  (3) at (-1, 0.5) {};
		\node  (4) at (-1.5, 1) {};
		\node  (5) at (-0.5, 1) {};
		\node  (21) at (-1, 0) {};
		\node  (45) at (1, 0.5) {};
		\node  (48) at (1, 0) {};
		\node  (66) at (-2, 0) {};
		\node  (67) at (8, 0) {};
		\node  (68) at (1, 0.5) {};
		\node  (69) at (0.5, 1) {};
		\node  (70) at (1.5, 1) {};
		\node  (71) at (3, 0.5) {};
		\node  (72) at (3, 0) {};
		\node  (73) at (3, 0.5) {};
		\node  (74) at (2.5, 1) {};
		\node  (75) at (3.5, 1) {};
		\node  (76) at (5, 0.5) {};
		\node  (77) at (5, 0) {};
		\node  (78) at (5, 0.5) {};
		\node  (79) at (4.5, 1) {};
		\node  (80) at (5.5, 1) {};
		\node  (81) at (7, 0.5) {};
		\node  (82) at (7, 0) {};
		\node  (83) at (7, 0.5) {};
		\node  (84) at (6.5, 1) {};
		\node  (85) at (7.5, 1) {};
		
		\node [red] at (-1,1.8) {\large $\Phi^* (x_1,t_1)$};
		\draw  node[fill= red,circle,scale=3,above] (86) at (-1, 0.63) {};
		\draw [green, ultra thick] (3.center) to (21.center) node[below]{\large $\Phi(r_1,t_1)$};
		\draw [green, ultra thick, bend right=90, looseness=1.75] (4.center) to (5.center);
		\draw [green, ultra thick] (45.center) to (48.center) node[below]{\large $\Phi(r_2,t_1)$};
		\draw [green, ultra thick][->] (66.center) to (67.center) ;
		\draw [green, ultra thick, bend right=90, looseness=1.75] (69.center) to (70.center);
		\draw [green, ultra thick] (71.center) to (72.center) node[below]{\large $\Phi(r_3,t_1)$};
		\draw [green, ultra thick, bend right=90, looseness=1.75] (74.center) to (75.center);
		\draw [green, ultra thick] (76.center) to (77.center) node[below]{\large $\Phi(r_4,t_1)$};
		\draw [green, ultra thick, bend right=90, looseness=1.75] (79.center) to (80.center);
		\draw [green, ultra thick] (81.center) to (82.center) node[below]{\large $\Phi(r_5,t_1)$};
		\draw [green, ultra thick, bend right=90, looseness=1.75] (84.center) to (85.center);
		\node[green] at (8.5,-0.2) {\large $... \,\,\,\Phi(r_n,t_1)$};
\end{tikzpicture}
\caption[Networking formalism diagram]{A diagram illustrating one of the possible pairings counted in the networking functional through the use of the fields $\Phi$ and $\Phi^*$. The configuration in the diagram depicts the motor protein (red) with position x(t) attaching to the binding site at position $r_1$.}
\label{fig:networkingDiagram}
\end{figure*}
The constraints in the networking functional need to be applied at each time step of the motor's motion. Consider the discretisation of $x(t)$ into a set of positions $x_j$  corresponding to discrete time steps $t_1,t_2,t_3,...$, such that for example $x_1 = x(t_1)$. One of the possible pairings, or attachments, is depicted in Fig. \ref{fig:networkingDiagram}. In this case the motor is attached to the first binding site, mathematically expressed as $\delta(r_1-x_1)$, at a time $t_1$. Using this discrete notation, the networking functional may now be applied at each time step $t_j$ by taking a product over $j$ as follows:
\begin{eqnarray}
\Pi_j Q(x_j,t_j) = \mathcal{N}_\Phi\int [ \mathrm{d}\Phi] [\mathrm{d} \Phi^*]\,\, \big\{\Pi_j \Pi_n (1+\Phi(r_n,t_j'))\nonumber\\
\times \Phi^*(x_j,t_j) 
 \mathrm{e}^{-\alpha \int_{y,t} \, \Phi(y,t)\,\Phi^*(y,t)} \big\}\, .   
\label{eq:networkingConstraintTimeProduct}
\end{eqnarray}

The networking functional may, once again, be rewritten such that the fields $\Phi$ and $\Phi^*$ appear in the exponent. Conveniently, doing so also allows the introduction of a binding site density
\begin{equation}
\rho(r,t) = \Sigma_n \delta(r-r_n(t))\,.
\label{eq:bindingsitedensityFormaldef}
\end{equation}
where the $r_n$ give the positions of each of the binding sites along the filament. Since the rest of the formalism is already dynamic, the binding site density may, at least formally, be defined as a time dependent quantity such that the positions of each of the binding sites can vary with time. With this, the binding site density may be formally introduced into the exponent as follows:
\begin{align}
\Pi_j\Pi_n(1+\Phi(r_n,t_j)) &{}= \mathrm{e} ^{\Sigma_j\Sigma_n\mathrm{ln}(1+\Phi(r_n,t_j))}\\
&{}= \mathrm{e} ^{\Sigma_j\int_{r}\rho(r,t_j)\mathrm{ln}(1+\Phi(r,t_j))}.
\label{eq:exponentiation}
\end{align}
Finally, taking the continuum limit of the time discretisation, the sums over $j$  become integrals over $t$ such that

\begin{multline}
Q[x(t)]=\mathcal{N}_\Phi \int [ \mathrm{d}\Phi] [\mathrm{d} \Phi^*] \ \mathrm{e}^{\frac{1}{\tau}\int_{r,t}\rho(r,t)\mathrm{ln}(1+\Phi(r,t))}\\
\times \mathrm{e}^{\frac{1}{\tau}\int_{t}\mathrm{ln}(\Phi^*(x(t),t))-\alpha \int_{y,t} \, \Phi(y,t)\,\Phi^*(y,t)}.
\label{eq:networkingconstraintFinal}
\end{multline}
Here $\tau$ gives the constant time interval separating the times $t_j$ in the discretisation of the networking functional and hence also the time interval at which the networking functional is applied to the motor protein. At times not equal to an increment of $\tau$, the motor is therefore not constrained to the network, but is able to diffuse freely subject only to the forces of the  coupled Langevin equations (eqs.~\eqref{eq:LangevinEqns}). Thus, this networking functional provides a formal expression requiring the periodic networking of the motor to one of a configuration of possible binding sites in a manner which is suited for combination with the Martin-Siggia-Rose formalism.

\section{Collective description}
\label{sec:collective}
Moving  towards a  collective description of  motor-driven transport, a concentration of $N$ motor proteins may be defined as follows:
\begin{equation}
    C(r,t) = \sum_{a=1}^{N} \delta (r - x_a(t)).
    \label{eq:concentrationDef}
\end{equation}
The system may conveniently be transformed to such collective field variables during the implementation of a random phase approximation (RPA).The RPA provides a method for approximating non-Gaussian path integrals with Gaussian path integrals. The approximation is valid for small fluctuations about the average of a collective variable. In this case, the variable under consideration is the position of the motor $x(t)$. This is of course a dynamical quantity and therefore the corresponding response field needs to form a part of this approximation. This is done in Appendix  \ref{app:rpa}, following a method that has been implemented to study collective dynamics of polymer solutions \cite{fredricksonCollectiveDynamicsPolymer1990} as well as particle diffusion in elastically coupled narrow channels \cite{Mateyisi2014} amongst others. 

The RPA is implemented within eq.~\eqref{eq:Zraw}, providing the generating functional of the system without the networking functional:
\begin{multline}
    Z_\mathrm{RPA} = \mathcal{N} \int [\mathrm{d} C_k] [\mathrm{d} \hat{C}_k]   \mathrm{e}^{\frac{i}{8\pi^2} \int_k \int_\omega  C_k(\omega) \mathcal{B}^{-1}_{-k}(-\omega) \hat{C}_{-k}(-\omega) }\\
    \times\,\mathrm{e}^{-\frac{1}{8 \pi^2} \int_k \int_\omega  \hat{C}_k(\omega) \mathcal{B}^{-1}_{k}(\omega) \mathcal{A}_{k}(\omega)\mathcal{B}^{-1}_{-k}(-\omega)  \hat{C}_{-k}(-\omega)}\,.
    \label{eq:ZRPAresult}
\end{multline}
This generating functional contains the dynamical behaviour of the motor and cargo coupling,  essentially representing a system without a filament configuration where the motors move freely in space. The effects of the networking of the motor to a filament may be incorporated by introducing the networking functional into eq.~\eqref{eq:ZRPAresult}. Therefore, the networking functional given by eq.~\eqref{eq:networkingconstraintFinal} needs to be rewritten to be consistent with the collective variable notation introduced in the RPA. For now, the networking functional will be rewritten to depend on $C(r,t,)$; temporarily deferring the calculation of the Fourier transform.

To rewrite the networking functional,  a product over $N$ motors is introduced as follows:
\begin{eqnarray}
 \Pi_ {a=1}^N Q[x_a(t)]=\mathcal{N}_\Phi \int [ \mathrm{d}\Phi] [\mathrm{d} \Phi^*] \  \big\{ \mathrm{e}^{\frac{N}{\tau}\int_{r,t}\rho(r,t)\mathrm{ln}(1+\Phi(r,t))} \nonumber\\
 \times \mathrm{e}^{\frac{1}{\tau}\int_{t}\sum_ {a=1}^N\mathrm{ln}(\Phi^*(x_a(t),t))- N\alpha \int_{y,t} \, \Phi(y,t)\,\Phi^*(y,t)}\big\}\,.\nonumber\\
\label{eq:networkingconstraintProduct}
\end{eqnarray}
Invoking  eq.~\eqref{eq:concentrationDef}, this leads to the dependence on the concentration of motors by the introduction of another spatial integral in the $x$-dependent term:
\begin{eqnarray}
 Q[C(r,t)]=\mathcal{N}_\Phi \int [ \mathrm{d}\Phi] [\mathrm{d} \Phi^*] \big\{\mathrm{e}^{\frac{N}{\tau}\int_{r,t}\rho(r,t)\mathrm{ln}(1+\Phi(r,t))} \nonumber \\
 \times \mathrm{e}^{\frac{1}{\tau}\int_{r,t}C(r,t)\mathrm{ln}(\Phi^*(r,t))- N\alpha \int_{y,t} \, \Phi(y,t)\,\Phi^*(y,t)}\big\}\,.\,\,\,\,\,\,
\label{eq:networkingconstraintConcentration}
\end{eqnarray}
To implement the non-Gaussian integrals over the fields that have been introduced in the networking functional in eq.~\eqref{eq:networkingconstraintConcentration}, a saddle point approximation may be utilised. In this approximation, the solution is found by expanding the argument of the exponent about its minimum value (see e.g. \cite{Mateyisi2011}). This is done in Appendix \ref{app:saddle}, yielding the  following expression for the networking functional:
\begin{eqnarray}
    Q[C(r,t)]= \mathcal{N}_\Phi \mathrm{e}^{ \frac{N}{\tau} \int_{r,t} \rho(r,t) \mathrm{ln}\left(\frac{N \rho(r,t) }{N \rho(r,t) -C(r,t)} \right)} \nonumber 
    \\ \times \mathrm{e}^{ \frac{1}{\tau}\int_{r,t}  C(r,t)\left(\mathrm{ln}\left(\frac{N \rho(r,t) -C(r,t)}{N\alpha \tau} \right)-1\right)}\, .
    \label{eq:Qsaddle}
\end{eqnarray}
What remains is to coherently combine eq.~\eqref{eq:Qsaddle} with eq.~\eqref{eq:ZRPAresult} and evaluate the integrals over the motor concentration and its corresponding response field.

\subsection{Small fluctuation expansions}
Following Ref.~\cite{dutoitDynamicalNetworkingUsing2025} and to simplify the dependence on the motor concentration in the networking functional, a Taylor expansion may be utilised in such a way that the approximation is along the same lines as that of the RPA. The reader may recall that within the RPA, terms that were merely dependent on the average motor concentration were neglected. This way the results obtained from the RPA highlight the fluctuations around the average motor concentration, which are assumed to be small. The same idea may be utilised again here. To clarify, note that the motor concentration may be written as the sum of the average motor concentration, say $\bar{C}$, and a fluctuating component $\Delta C(r,t)$, i.e.
\begin{equation}
    C(r,t) = \bar{C} + \Delta C(r,t)\, .
\end{equation}
Thus the Taylor expansion needs to be valid where $\Delta C(r,t) \rightarrow 0$. In addition, the resulting terms that depend only on the average motor concentration $\bar{C}$ may be neglected. Here, this expansion will be applied to the argument of the exponent of eq.~\eqref{eq:Qsaddle}. Up to second order in $C(r,t)$, this yields the following approximation to eq.~\eqref{eq:Qsaddle}:

\begin{equation}
    Q[C(r,t)]= \mathcal{N}_\Phi \mathrm{e}^{-\frac{1}{\tau}\int_{r,t} \mathrm{ln}(\frac{\alpha\tau}{\rho(r,t)}) C(r,t) - \int_{r,t}\frac{1}{2N \tau \rho(r,t) }(C(r,t))^2}\, . 
    \label{eq:QafterCExpansion}
\end{equation}
The dependence on the motor concentration is now in a form that allows easy functional integration, but the dependence on the binding site density $\rho(r,t)$  remains problematic when trying to implement the spatial and temporal Fourier transforms. There are various ways in which this dependence on $\rho(r,t)$ in eq.~\eqref{eq:QafterCExpansion} may be dealt with, depending on which assumptions one chooses to make about the distribution of the sites in the filament network.  To illustrate this, two scenarios will be considered.

The binding site density may unsurprisingly be written as the sum of the average binding site density $\bar{\rho}$ and a fluctuating component $\Delta \rho(r,t)$, such that the networking functional becomes
\begin{eqnarray}
      Q[C(r,t)]= \mathcal{N}_\Phi \mathrm{e}^{-\frac{1}{\tau}\int_{r,t} \mathrm{ln}(\frac{\alpha\tau}{ \bar{\rho} +\Delta\rho(r,t)}) C(r,t) }\nonumber\\
      \times \mathrm{e}^{- \int_{r,t}\frac{1}{2N \tau ( \bar{\rho} +\Delta\rho(r,t)) }(C(r,t))^2}\, .
     \label{eq:QrhoFluctuating}
\end{eqnarray}
If one assumes, that the binding sites are uniformly distributed such that $\rho(r,t)= \bar{\rho}$ or equivalently $\Delta \rho(r,t) = 0$, the expression is greatly simplified such that the spatial and temporal Fourier transforms may be calculated directly to obtain
\begin{equation}
    Q[C_k(\omega)]_{\bar{\rho}} = \mathcal{N}_\Phi \mathrm{e}^{ - \frac{1}{8 \pi ^2N \bar{\rho}L \tau}\int_{k,\omega}C_k(\omega)C_{-k}(-\omega)}\, .
    \label{eq:QrhoAve} 
\end{equation}
Here $L$ gives the length over which the binding sites are distributed. Note that one of the terms has been discarded, since the Fourier transform yields $C_0$ which is the average previously denoted by $\bar{C}$. The expression above gives the networking functional for any network with uniformly distributed binding sites, which will from here onward be referred to as a \textit{homogeneous network}.

The goal here is to develop a formalism that may be applied to transport on a variety of network configurations. Therefore, considering only networks with such specific properties as uniformly distributed binding sites is not ideal. To obtain a similar expression for the networking functional that is applicable in a more general scenario, a small fluctuation expansion may be implemented for the binding site density as well. For this, expand the argument of the exponent of eq.~\eqref{eq:QrhoFluctuating} around $ \Delta \rho \rightarrow 0$ up to first order to obtain
\begin{eqnarray}
      Q[C(r,t)]= \mathcal{N}_\Phi \mathrm{e}^{\int_{r,t}\left( \frac{\Delta\rho(r)}{\bar{\rho}}-\mathrm{ln}(\frac{\alpha\tau}{\bar{\rho}})\right) C(r,t) } \nonumber \\
      \times \mathrm{e}^{- \int_{r,t}\left( \frac{\Delta \rho(r)}{2 N \bar{\rho}^2}- \frac{1}{2 N \bar{\rho}}\right)(C(r,t))^2}\, . 
     \label{eq:QrhoNonUniform}
\end{eqnarray}
This expansion has been executed up to first order only, since the saddle point approximation was also only done to first order for the sake of mathematical simplicity. The spatial and temporal Fourier transforms, may now be taken such that
\begin{multline}
      Q[C_k(\omega)]= \mathcal{N}_\Phi \mathrm{e}^{\frac{1}{4 \pi^2}\int_{k,\omega}\frac{1}{\bar{\rho} \tau} \Delta\rho_{k}(\omega)C_{-k}(-\omega)} \\
      \times \mathrm{e}^{  - \frac{1}{4 \pi^2}\int_{k,\omega}\frac{1}{2N \bar{\rho} \tau}C_k(\omega)C_{-k}(-\omega)}\\
      \times \mathrm{e}^{ \frac{1}{4 \pi^2}\int_{k,\omega}\int_{k',\omega'}\frac{1}{2N \bar{\rho}^2 \tau}C_k(\omega)\Delta \rho_{-k-k'}(-\omega-\omega') C_{k'}(\omega')}\, . 
     \label{eq:QrhoNonUniformFT}
\end{multline}
Evidently this expression is more complicated than the scenario for uniformly distributed binding sites, but remains useful since it provides the opportunity for considering a variety of binding site densities. This will be addressed in more detail in the next section.

\section{Results}
\label{sec:results}

The full generating functional for motors on a network may be obtained by introducing the relevant networking functional into the generating functional obtained from the RPA (eq.~\eqref{eq:ZRPA}). Noting that the networking functional does not depend on $\hat{C}_k$, the Gaussian integral over $\hat{C}_k$ may be implemented at this stage of the calculation, leaving the following generating functional:
    \begin{eqnarray}
    Z[J_k(\omega)] = \mathcal{N} \int [\mathrm{d} C_k] \,\, \big\{ Q[C_k(\omega)]  \mathrm{e}^{+\frac{1}{4\pi^2} \int_{k,\omega}  J_k(\omega) C_{-k}(-\omega)} \nonumber \\
      \times \mathrm{e}^{-\frac{1}{8\pi^2} \int_{k,\omega}  C_k(\omega) \mathcal{A}^{-1}_{k}(\omega)   C_{-k}(-\omega) }\big\}\,.\nonumber\\ 
       \label{eq:ZfullQgeneral}
    \end{eqnarray} 
A source term, $J_k(\omega)$, for the motor protein concentration has been introduced above. The preceding discussions have revealed that the final form of the networking functional depends on the assumptions that may be made about the filament network. Thus, to extract further meaning from the eq.~\eqref{eq:ZfullQgeneral}, the scenarios of homogeneous and non-homogeneous networks need to be considered separately. 

\subsection{Transport along  homogeneous networks}

Substituting eq. \eqref{eq:QrhoAve} into eq.~\eqref{eq:ZfullQgeneral} and executing the Gaussian path integral over the concentration of motor proteins yields 
\begin{equation}
\begin{aligned}
Z[J_k(\omega),\bar{\rho}] = \mathcal{N} \mathrm{e}^{\frac{1}{8\pi^2} \int_{k,\omega}  J_k(\omega)\left(\frac{A_k(\omega)}{1 +  N \bar{\rho} L \tau A_k(\omega) }\right)J_{-k}(-\omega)}\,.
\end{aligned}
\end{equation}
This is the final version of the generating functional for uniformly distributed binding sites. Recalling that partial derivatives of the generating functional may be used according to eq.~\eqref{eq:aveFormula}, the correlation function for the concentration of motor proteins amounts to
\begin{equation}
    \langle C_k(\omega)C_{-k}(-\omega) \rangle=\frac{2Dk^2}{D^2 k^4 + \frac{2Dk^2}{N \bar{\rho} L \tau}+ (\omega - k v_\mathrm{drift})^2}
    \label{eq:CorrRhoAve}
\end{equation}
where $D$ is the diffusion coefficient given by 
\begin{multline}
 D =\frac{ \omega^2}{2}\left[ \frac{\left(-\frac{\kappa ^2 (\gamma_{R} \omega -\mathrm{i} \kappa )}{\gamma_R^2 \omega ^2+2 \mathrm{i} \gamma_R \kappa  \omega +\kappa ^2}+\gamma_x \omega -i \kappa \right)^2}{ \left(\lambda_x+\frac{\kappa ^2 \lambda_R}{\left(\gamma_R^2 \omega ^2+2 \mathrm{i} \gamma_R \kappa  \omega +\kappa ^2\right)^2}\right)}\right.\\
 \left..+\frac{\kappa ^2 (\gamma_R \omega -\mathrm{i} \kappa )^2}{ \lambda_R \left(\gamma_R^2 \omega ^2+2 \mathrm{i} \gamma_R \kappa  \omega +\kappa ^2\right)}+\frac{\kappa ^2}{ \lambda_R}\right]^{-1} \,.
 \label{eq:DiffCoeff}
\end{multline}

As a first investigation of this result, it is useful to compare this diffusion coefficient with the known diffusion coefficient of Brownian particles. Decoupling the motor proteins and the cargoes, eq.~\eqref{eq:DiffCoeff} should simply describe $N$ Brownian particles. For $\kappa = 0$ the diffusion coefficient(eq.~\eqref{eq:DiffCoeff}) simplifies  to
\begin{equation}
D_{\kappa = 0} =\frac{ \lambda_x}{2 \gamma_x^2} = \frac{k_\mathrm{B} T} {\gamma_x} \,,
\label{eq:Dkapp0}
\end{equation}
which corresponds exactly to what is expected. 

The interpretation of the quantity $\bar{\rho}\tau$, appearing in eq.~\eqref{eq:CorrRhoAve}, may perhaps be more suitably illustrated with a specific scenario of binding site densities in mind. A straight filament with uniformly distributed binding sites, may for example have $\bar{\rho} = \frac{1}{\ell}$ where $l$ is the distance between neighbouring binding sites. In this case, the relationship of $(\bar{\rho}\tau)^{-1} =\frac{\ell}{\tau}$  very clearly has the dimensions of a speed. Since $\tau$ was introduced as a time constant pertaining to the intervals at which the networking functional is applied and $\ell$ is the distance between neighbouring binding sites, this speed is in fact the speed at which the motor \textit{hops} from one binding site to the next. One might therefore call $(\bar{\rho}\tau)^{-1}$ the hopping speed of a motor protein that is progressing along the binding sites. 

The hopping speed, together with the diffusion coefficient appearing in the correlation function (eq.~\eqref{eq:CorrRhoAve}), reveals a modified effective diffusion similar to the Brownian motion example presented in the original development of this dynamical formalism~\cite{dutoitDynamicalNetworkingUsing2025}. In the present context,
the motors may be thought of as freely diffusing particles which only attach to the binding sites at a specified time interval $\tau$. 
The diffusion coefficient therefore incorporates the additional drag arising from the attached cargo, while the hopping speed sets the rate at which motors progress between neighbouring binding sites.
\subsection{Transport along non-homogeneous networks}
\label{sec:non-hom}
As before, substituting the relevant networking functional, in this case eq.~\eqref{eq:QrhoNonUniformFT}, into eq.~\eqref{eq:ZfullQgeneral} yields the generating functional:
\begin{widetext}
\begin{eqnarray}
    Z[J_k(\omega), \bar{\rho}, \Delta \rho_k (\omega)] = \mathcal{N}\mathcal{N}_\Phi \int [\mathrm{d} C_k] \,\, \mathrm{e}^{\frac{1}{4 \pi^2}\int_{k,\omega}\left(\frac{1}{\bar{\rho} \tau}\Delta \rho_{k}(\omega) + J_k(\omega) \right)C_{-k}(-\omega)} \nonumber\\
      \times \mathrm{e}^{ -\frac{1}{8 \pi^2}\int_{k,\omega}\int_{k',\omega'} C_k(\omega)\left(\delta(\omega+ \omega')\delta(k+k')(\frac{1}{N \bar{\rho}\tau}+  \frac{1}{4}\mathcal{A}^{-1}_{k}(\omega))-\frac{1}{N \bar{\rho}^2\tau}\Delta \rho_{-k-k'}(-\omega-\omega') \right) C_{k'}(\omega')}\,. 
    \label{eq:ZfullNonHom}
\end{eqnarray}
\end{widetext}
Due to the asymmetry in both $\omega$ and $k$ in the exponents, the Gaussian path integral over the concentration of motor proteins is not as easily executable as before. Previously, the symmetry of the second order term in the exponent allowed one to easily determine the inverse of its coefficient. To extract the results for a specific configuration of non-uniformly distributed binding sites, the simplest solution may be to perform this calculation numerically. The power of the formalism, however, does not lie in applications to such specific scenarios. 

The generating functional eq.~\eqref{eq:ZfullNonHom}, already contains a lot of information of a rather complicated system, but has by no means reached a final form. The possibility still remains to average over a set of binding site densities. To do this, one would have to introduce some probability distribution for the binding site density into the generating functional and execute another path integral. Due to the significance of the quantity $(\bar{\rho}\tau)^{-1}$, as a hopping speed, the suggestion here would be to choose some suitable value for $\bar{\rho}$ and introduce a probability distribution $P[\Delta \rho_k(\omega)]$ for the fluctuations around this value, as follows:
\begin{widetext}
\begin{eqnarray}
    \langle Z[J_k(\omega), \bar{\rho}]  \rangle_\Delta \rho = \mathcal{N}\mathcal{N}_\Phi \int [\mathrm{d} C_k][\mathrm{d} \Delta \rho_k] \,\, P[\Delta \rho_k(\omega)] \mathrm{e}^{\frac{1}{4 \pi^2}\int_{k,\omega}\left(\frac{1}{\bar{\rho} \tau}\Delta \rho_{k}(\omega) + J_k(\omega) \right)C_{-k}(-\omega)} \nonumber\\
      \times \mathrm{e}^{ -\frac{1}{8 \pi^2}\int_{k,\omega}\int_{k',\omega'} C_k(\omega)\left(\delta(\omega+ \omega')\delta(k+k')(\frac{1}{N \bar{\rho}\tau}+  \frac{1}{4}\mathcal{A}^{-1}_{k}(\omega))-\frac{1}{N \bar{\rho}^2\tau}\Delta \rho_{-k-k'}(-\omega-\omega') \right) C_{k'}(\omega')}\,. 
    \label{eq:ZfullNonHomNetworkAve}
\end{eqnarray}
\end{widetext}
Depending on the probability distribution that is introduced, this average over the fluctuating component of the binding site density could be either quenched or annealed. Should the probability distribution have only a spatial dependence, but no time dependence, the average will be quenched. This scenario would model the diffusion of motor proteins on a configuration of binding sites under the assumption that the configuration of binding sites does not change at the time scales at which the motors diffuse. Such an average would, for example, allow one to take into account the typical spatial variations in the density of cytoskeleton filaments for various regions of a cell (see e.g. \cite{Azote2019}). Alternatively, the  possibility remains to introduce a probability distribution with some or other time dependence and obtain an annealed average instead.

\section{Conclusions}
\label{sec:Conclusions}

This paper has built upon the dynamical networking framework of Ref.~\cite{dutoitDynamicalNetworkingUsing2025}, adapting it to the setting of motor-driven transport along cytoskeletal filaments. In this formulation, the networking functional is utilised along with a set of Langevin equations in a Martin-Siggia Rose representation. The networking theory performs the role of periodically constraining the position of the motor protein to one of a set of possible attachment points along a filament. After implementing a saddle point approximation and a random phase approximation, a correlation function is obtained for a collection of motor proteins diffusing on a filament with homogeneously distributed attachment points.  This result reveals the diffusion coefficient of the motor proteins, which simplifies to that of simple Brownian particles when decoupling the cargoes from the motors, as one would expect. An additional term appears in the correlation function, arising from the networking theory. This term reveals the \textit{hopping speed} of the motor proteins on the binding sites. This hopping speed provides scale-able parameters for the length separation between neighbouring binding sites and the time interval at which the motor protein is required to be instantaneously constrained to one of these sites. Thus, the networking functional was implemented successfully in the dynamical system in such a manner that control may be kept over the manner in which it is implemented so as not to introduce unwanted effects into the system. 

In order to illustrate how this formalism may be applied to a variety of filament configurations, including networks of branched and intersecting filaments, the scenario of motor-driven transport along filaments with non-homogeneously distributed binding sites is presented in Section \ref{sec:non-hom}. The formalism is shown to allow for quenched and in principle even annealed averaging over an ensemble of binding site densities. Applying this in a manner that accounts for various cytoskeleton densities would allow one to investigate how various properties of the cytoskeleton might affect the diffusion of motor proteins within a cell. This could include spatial variations in cytoskeleton networks, or potentially dynamic fluctuations in the network. The model further allows for the inclusion of physical phenomena entirely unrelated to the cytoskeleton, such as excluded volume effects and the hydrodynamic coupling of motors to one another \cite[see e.g., ][]{Guerin2010}. 

Taken together, these results demonstrate how the dynamical networking framework can be extended beyond polymers to active processes, such as intracellular transport and collective motor dynamics, providing a flexible basis for exploring more complex scenarios and applications.

\begin{acknowledgments}
This work is based on the research supported in part by the National Research Foundation of South Africa (Grant Numbers 99116, MND210620613719 and PMDS240820261081) and the National Institute of Theoretical and Computational Sciences (Grant Number MND190813466136).\\

\textbf{Author contribution}: Conceptualisation (NdT and KKMN), Calculations (detailed calculations and numerical work: NdT; approximation approach, checks: NdT and KKMN), Interpretation (NdT and KKMN), Writing (NdT), Editing (NdT and KKMN)
\end{acknowledgments}

\appendix

\section{\label{app:rpa}The Random Phase Approximation (RPA)}
The RPA calculation presented below follows closely the derivation given in the appendix of Ref.~\cite{dutoitDynamicalNetworkingUsing2025}, with adjustments made to match the notation and the specific context of motor-driven transport considered here.\\

The integrals over the variables other than $x$ and $\hat{x}$ in eq.~\eqref{eq:Zraw} may be evaluated such that the generating functional is of the form:
\begin{equation}
Z =  \mathcal{N} \int [\mathrm{d} x(t)]  [\mathrm{d} \hat{x}(t)] \mathrm{e}^{\mathcal{F}[x, \hat{x}]}.
\label{eq:ZrpaStart}
\end{equation}
These calculations are straightforward, since the path integrals over the $f_R(t)$, $f_x(t)$ and $\hat{R}(t)$ are Gaussian and the integral over $R(t)$ becomes Gaussian when Fourier transformed to frequency space. 

Gradually moving towards an expression that depends on a concentration of motor proteins, $N$ motor proteins are now considered --- each of which has a generating functional of the form of eq.~\eqref{eq:ZrpaStart}. The product of these, produces a generating functional for all $N$ motors:
\begin{multline}
Z =  \mathcal{N}\int [\mathrm{d} x_1(t)] [\mathrm{d} x_2(t)]...[\mathrm{d} x_N(t)] [\mathrm{d} \hat{x}_1(t)][\mathrm{d} \hat{x}_2(t)]\\
...[\mathrm{d} \hat{x}_N(t)] \mathrm{e}^{\sum_{a=1}^{N} \mathcal{F}[x_a, \hat{x}_a]}.
\label{eq:ZrpaStartN}
\end{multline}
For these $N$ motor proteins, a concentration is defined according to eq.~\eqref{eq:concentrationDef}. Following Refs.~\cite{fredricksonCollectiveDynamicsPolymer1990} and \cite{dutoitDynamicalNetworkingUsing2025}, however, it is mathematically more convenient to utilise the spatial Fourier transform, $C_k$, of the concentration along with its corresponding auxiliary variable $\hat{C}_k$.
These variables are incorporated by multiplying the following into eq.~\eqref{eq:ZrpaStartN}
\begin{eqnarray}
    \int [\mathrm{d} C_k] [\mathrm{d} \hat{C}_k] \big\{ \delta(C_k - \sum_{a=1}^{N}\mathrm{e}^{i k x_a(t)})\nonumber\\
    \times \delta (\hat{C}_k - i\sum_{a=1}^{N}k x_a(t)\mathrm{e}^{i k x_a(t)}) \big\}\, .
\end{eqnarray}
    This is equivalent to
    \begin{multline}
\mathcal{N} \int [\mathrm{d} C_k] [\mathrm{d} \hat{C}_k] [\mathrm{d} \psi_k] [\mathrm{d} \hat{\psi}_k]\mathrm{e}^{i \int_{k,t}\psi_k(C_k - \sum_{a=1}^{N}\mathrm{e}^{i k x_a(t)})}\\
    \times \mathrm{e}^{i \int_{k,t}\psi_k(\hat{C}_k - i\sum_{a=1}^{N}k x_a(t)\mathrm{e}^{i k x_a(t)})}.
    \end{multline}
Further utilising some second order expansions, the generating functional becomes:
\begin{multline}
    Z = \mathcal{N}\int [\mathrm{d} C_k] [\mathrm{d} \hat{C}_k] [\mathrm{d} \psi_k] [\mathrm{d} \hat{\psi}_k]\mathrm{e}^{i \int_{k,t}\psi_k C_k +i \int_{k,t}\psi_k \hat{C}_k} \\
    \times \Bigg\{  \int [\mathrm{d} x_1(t)] [\mathrm{d} x_2(t)]...[\mathrm{d} x_N(t)] [\mathrm{d} \hat{x}_1(t)][\mathrm{d} \hat{x}_2(t)]...[\mathrm{d} \hat{x}_N(t)]\\
    \times  \mathrm{e}^{\sum_{a=1}^{N} \mathcal{F}[x_a, \hat{x}_a]} \left(1-  i \int_{k,t}\psi_k \sum_{a=1}^{N} \mathrm{e}^{i k x_a(t)}\right.\\ \left.+ \int_{k,t}\hat{\psi}_k \sum_{a=1}^{N} k \hat{x}_a(t)\mathrm{e}^{i k x_a(t)} \right. \\ \left.- i \int_{k,t}\int_{k',t'} \psi_k \hat{\psi}_{k'} \sum_{a=1}^{N} \sum_{\alpha=1}^{N} k' \hat{x}_\alpha (t)\mathrm{e}^{i k x_a(t) + i k' x_\alpha(t')} \right. \\ \left.+\frac{1}{2} \int_{k,t}\int_{k',t'} \hat{\psi}_k \hat{\psi}_{k'} \sum_{a=1}^{N} \sum_{\alpha=1}^{N} k k' \hat{x}_a (t) \hat{x}_\alpha (t')\mathrm{e}^{i k x_a(t) + i k' x_\alpha(t')} 
    \right. \\ 
    \left.+\frac{1}{2} \int_{k,t}\int_{k',t'} \psi_k \psi_{k'} \sum_{a=1}^{N} \sum_{\alpha=1}^{N} \mathrm{e}^{i k x_a(t) + i k' x_\alpha(t')}\right)\Bigg\}.
\end{multline}
At this point, the functional integrals over the $x_a(t)$ and $\hat{x}_a(t)$ are evaluated for each of the terms in the expansion. 
One of these functional integrals merely results in the average concentration of motor proteins. The crux of this approximation is that this average concentration, or $k=0$, term may be omitted.
After some final mathematical manipulation of the remaining functional integrals, one obtains:
\begin{multline}
    Z = \mathcal{N}\int [\mathrm{d} C_k] [\mathrm{d} \hat{C}_k] [\mathrm{d} \psi_k] [\mathrm{d} \hat{\psi}_k]\Bigg\{\mathrm{e}^{\frac{i}{4\pi^2} \int_k \int_\omega\psi_k C_k} \\
        \times \mathrm{e}^{\frac{i}{4\pi^2} \int_k \int_\omega \psi_k \hat{C}_k} \left(1- \frac{1}{8 \pi^2} \int_k \int_\omega  \psi_k(\omega) \mathcal{A}_{k}(\omega) \psi_{-k}(-\omega)\right.\\ \left. +\frac{i}{4\pi^2} \int_k \int_\omega  \psi_k(\omega) \mathcal{B}_{-k}(-\omega) \hat{\psi}_{-k}(-\omega) \right)\Bigg\}
\end{multline}
where 

\begin{equation}
    \begin{aligned}
 \mathcal{A}_{k}(\omega) =\frac{2 D  k^2}{D  ^2k^4
 +\left(\omega - k v_\mathrm{drift}  \right)^2}
 \label{eq:Akdef}
\end{aligned}
\end{equation}
and
\begin{multline}
    \mathcal{B}_{-k}(-\omega) =   k^{N+1}\mathcal{A}_{-k}(-\omega) \mathrm{e}^{-k^2 D} \\
    \times \frac{\left(-\frac{\kappa ^2 (\gamma_{R} \omega -\mathrm{i} \kappa )}{\gamma_R^2 \omega ^2+2 \mathrm{i} \gamma_R \kappa  \omega +\kappa ^2}+\gamma_x \omega -i \kappa \right)}{\left(\lambda_x+\frac{\kappa ^2 \lambda_R}{\left(\gamma_R^2 \omega ^2+2 \mathrm{i} \gamma_R \kappa  \omega +\kappa ^2\right)^2}\right)} \\
    \times\left( \sqrt{ \frac{2 \pi \left(\lambda_x+\frac{\kappa ^2 \lambda_R}{\left(\gamma_R^2 \omega ^2+2 \mathrm{i} \gamma_R \kappa  \omega +\kappa ^2\right)^2}\right) }{\left(-\frac{\kappa ^2 (\gamma_{R} \omega -\mathrm{i} \kappa )}{\gamma_R^2 \omega ^2+2 \mathrm{i} \gamma_R \kappa  \omega +\kappa ^2}+\gamma_x \omega -i \kappa \right)^3}}\right)^N\\
     \, .
\end{multline}

What remains, is to move everything back into the exponent and implement a few Gaussian integrals. This exercise delivers the following result for the generating functional:
\begin{multline}
    Z_\mathrm{RPA} = \mathcal{N} \int [\mathrm{d} C_k] [\mathrm{d} \hat{C}_k]   \mathrm{e}^{\frac{i}{4\pi^2} \int_k \int_\omega  C_k(\omega) \mathcal{B}^{-1}_{-k}(-\omega) \hat{C}_{-k}(-\omega) }\\
    \times\,\mathrm{e}^{-\frac{1}{8 \pi^2} \int_k \int_\omega  \hat{C}_k(\omega) \mathcal{B}^{-1}_{k}(\omega) \mathcal{A}_{k}(\omega)\mathcal{B}^{-1}_{-k}(-\omega)  \hat{C}_{-k}(-\omega)}, 
    \label{eq:ZRPA}
\end{multline}
with corresponding structure factor:
\begin{equation}
    \langle C_k(\omega) C_{-k}(-\omega)\rangle = \mathcal{A}_{k}(\omega).
\end{equation}

From the resulting correlation function of the RPA, we may extract the diffusion coefficient for  $N$ motor proteins each transporting a cargo as they move freely in space. If one were to consider the Fokker-Planck equation for the concentration of motor proteins $C_k(\omega)$, the diffusion coefficient corresponds to the coefficient of the $k^2$ term in the correlation of $C_k(\omega)$. Thus, eq.~\eqref{eq:DiffCoeff} turns out to be the diffusion coefficient for $N$ motor proteins and their cargoes that are moving freely in space.  

\section{\label{app:saddle}The saddle point approximation}
Let, $\mathcal{F}_Q$ denote the argument of the exponent in eq.~\eqref{eq:networkingconstraintConcentration}, i.e. 
\begin{multline}
\mathcal{F}_Q[\Phi,\Phi^*] =\frac{N}{\tau}\int_{r,t}\rho(r,t)\mathrm{ln}(1+\Phi(r,t))\\+\frac{1}{\tau}\int_{r,t}C(r,t)\mathrm{ln}(\Phi^*(r,t))- N\alpha \int_{y,t} \Phi(y,t)\Phi^*(y,t) .
\end{multline}
To find the saddle point the following expressions should thus be solved
\begin{subequations}
\begin{eqnarray}
0=\left.\frac{\partial \mathcal{F}_Q}{\partial \Phi(r,t)} \right|_{\bar{\Phi}^* ,\bar{\Phi}}\\
0=\left.\frac{\partial \mathcal{F}_Q}{\partial \Phi^*(r,t)} \right|_{\bar{\Phi}^* ,\bar{\Phi}}\\
\end{eqnarray}
\end{subequations}
leading to
\begin{subequations}
\begin{eqnarray}
\bar{\Phi}^*(r,t) = \frac{\rho(r,t)}{\alpha \tau (1+\bar{\Phi}(r,t))} \,,\label{eq:saddleSol1}\\ 
\bar{\Phi}(r,t)  = \frac{C(r,t)}{\tau N \alpha \bar{\Phi}^*(r,t)}\,. \label{eq:saddleSol2}\\
\end{eqnarray}
Simultaneously solving eqs.~\eqref{eq:saddleSol1} and \eqref{eq:saddleSol2}  yields
\begin{eqnarray}
\bar{\Phi}^*(r,t) = \frac{\rho(r,t)}{\tau \alpha}-\frac{C(r,t)}{N\tau \alpha}   \label{eq:saddleSolFinal1}\\
\bar{\Phi}(r,t) = \frac{C(r,t)}{N\rho(r,t)- C(r,t)} \,. \label{eq:saddleSolFinal2}
\end{eqnarray}
\end{subequations}
Merely substituting eqs.~\eqref{eq:saddleSolFinal1} and \eqref{eq:saddleSolFinal2} into 
\begin{equation}
   Q[C(r,t)]=\mathcal{N}_\Phi \mathrm{e}^{\mathcal{F}_Q [\bar{\Phi},\bar{\Phi}^*]}
\end{equation}
amounts to a first order expansion around the minimum of $\mathcal{F}_Q$, revealing the  result of the saddle point approximation.

\bibliographystyle{apsrev4-2} 
\bibliography{references}

\end{document}